\documentclass[aps,prd,preprint]{revtex4}

\usepackage{graphicx}
\usepackage{psfrag}

\begin{document}
\title{Uncertainties in Estimating $B_c$ Hadronic Production
and Comparisons of the Production at TEVATRON and LHC}
\author{Chao-Hsi Chang$^{1,2}$ \footnote{Email:
zhangzx@itp.ac.cn} and Xing-Gang Wu$^{2,3}$\footnote{Email:
wuxg@mail.ihep.ac.cn}}
\address{$^1$CCAST
(World Laboratory), P.O.Box 8730, Beijing 100080,
China.\footnote{Not correspondence address.}\\
$^2$Institute of Theoretical Physics, Chinese Academy of Sciences,
P.O.Box 2735, Beijing 100080, China.\\$^3$Institute of High Energy
Physics, P.O.Box 918(4), Beijing 100039, China}

\begin{abstract}
The uncertainties in estimating the hadronic production of the
$B_c$ meson are studied under the framework of the complete
$\alpha_s^4$ approach of the perturbative QCD and the gluon-gluon
fusion mechanism. Quantitative comparisons on the production at
TEVATRON and LHC are made. Considering the detectors at TEVATRON
and LHC, we have also estimated the production with proper
kinematic cuts. Based on the results, we conclude that the
experimental studies of the $B_c$ meson at the two colliders will
be complimentary and stimulative. We find that as c.m. energy is
increasing from RUN-I to RUN-II at TEVATRON, the
production cross-section increases about $20\%$.\\


\end{abstract}

\maketitle

\section{Introduction}

$B_c$-physics is attracting more and wide interests due to
experimental progresses \cite{lep,CDF,CDF1} and theoretical ones
\cite{B-work,mord,qigg,latt,prod0,prod,prod1,prod2,
prod3,avaa,prod4,spec,chen,dec,dec1,gem,life,MG,CCWZ} (especially
the experimental discovery \cite{CDF1}). Since the production
cross section is relatively small in comparison with the
production of the other heavy mesons ($B$, $D$ etc) and heavy
quarkonia ($J/\psi$, $\Upsilon$ etc), only at high energy hadronic
colliders with high luminosity, can one collect enough $B_c$
events for experimental studies
\cite{B-work,mord,qigg,prod0,prod,prod1,prod2,prod3}. Considering
the potential applications to the experimental feasibility studies
on the topic, a computer program for hadronic generating the $B_c$
meson, BCVEGPY, has been completed in Ref.\cite{wuxglun}, which
has been written in Fortran and can be conveniently implemented
into PYTHIA\cite{pythia}.

The hadronic production of the $B_c$-states with different spin
has been estimated by a lot of authors, and all are mainly based
on the gluon-gluon fusion mechanism, i.e. taking the hard
subprocess $gg\to B_c+b+\bar{c}$ as dominant one. The authors of
\cite{prod,prod4} gave the estimate in terms of the `fragmentation
approach', and the authors of \cite{prod1,prod2,prod3} did it in
the so-called `complete approach'. The results of the two
approaches agree essentially within theoretical uncertainties,
especially, at high $B_c$ transverse momentum $P_T$ \cite{prod2}
and when the fragmentation approach additionally takes into
account the contributions from gluon fragmentation \cite{prod4}.
The fragmentation approach is appropriate if one is only
interested in the information of the produced $B_c$ itself, and
its accuracy is improving with the increasing of $P_T$
\cite{prod2}. Comparatively, the advantage of the complete
approach is that it can also retain the information of the
accompanied $\bar{c}$ and $b$ quarks (jets). For experiments, to
retain more information of the events is more relevant. Therefore,
in BCVEGPY \cite{wuxglun}, the complete calculation approach has
been adopted.

Moreover according to perturbative QCD (pQCD), in addition to the
gluon-gluon fusion mechanism, there are several different
mechanisms for the production, such as that via the
quark-antiquark annihilation subprocess $q \bar{q}\to
B_c+b+\bar{c}$, and that via the color octet mechanism and {\it
etc.}. In Ref.\cite{prod1}, it is shown that the contributions to
the production from quark-antiquark annihilation are much smaller
than those from gluon-gluon fusion. It is mainly because of the
fact that the relevant luminosity for gluon-gluon fusion is a
product of the parton distribution functions (PDFs) for gluon
components in the colliding hadrons, whereas 6that for
quark-antiquark annihilation is a product of the quark and
anti-quark distribution functions respectively while in a high
energy collider, such as TEVATRON ($p\bar{p}$ collision $\sqrt
S=2$ TeV) at Fermilab and LHC ($pp$ collision $\sqrt S=14$ TeV) at
CERN, the luminosity for gluon-gluon fusion due to parton
distribution functions (PDFs) at the most kinematic region is much
higher than that for quark-antiquark annihilation \cite{6lcteq}.
Furthermore, in the annihilation subprocess, there is an
additional $S$-channel suppression from the virtual gluon
propagator, which will also make contributions to the production
cross section small. In the subsection II-C we will make a brief
comparison between those two mechanisms. According to
non-relativistic QCD (NRQCD) \cite{nrqcd}, for a double heavy
meson, such as $J/\psi, \eta_c, B_c,\cdots$, there may be
additional color-octet components which may contribute the
production substantially. The magnitude of the color-octet
components may be estimated with the NRQCD velocity scaling rule
($v\simeq \alpha_s(m_{B_c})\sim 0.1$). One may find that there is
no enhancement in the hadronic production (in contract to  of the
hidden flavored double heavy objects such as $J/\psi, \psi'$the
production which can gain one order or more in $\alpha_s$ via the
color octet components) and the color octet components in $B_c$
are definitely smaller than the color singlet ones (with the same
input parameters, the total contributions to the hadronic
production is $\lesssim v \sim v^2$ of the color-singlet one),
therefore, the contributions from color octet components of $B_c$
are ignorable at leading order.

The contributions to the production from the color octet and those
from the quark-antiquark annihilation mechanism were neglected in
Refs.\cite{prod,prod2,prod3,avaa,prod4}. In BCVEGPY \cite{wuxglun}
and thus in the present paper, the mechanisms via color-octet for
the $B_c$ production are ignored.

Although the estimated values for the hadronic production of $B_c$
in literature \cite{prod,prod2,prod3,prod4} are consist within the
theoretical uncertainties, in a few cases the differences on the
production can reach up to almost one magnitude order. Moreover,
RUN-II of TEVATRON is taking data, LHC is under constructing, and
various experimental feasibility studies of $B_c$ are in progress,
thus we think to know the theoretical uncertainties quantitatively
in estimates of the $B_c$ production, in addition, the precise
comparisons of the production at LHC and TEVATRON are interesting.

The newly developed generator BCVEGPY has been well tested by
carefully comparing the results with those in earlier references
\cite{prod1}. It, having been implemented into PYTHIA, can be
conveniently applied to simulate $B_c$ events (Monte Carlo
simulation). In the paper, we highlight the uncertainties of the
dominant gluon-gluon fusion mechanism, make a brief comparison on
the gluon-gluon fusion mechanism and the quark-antiquark
annihilation mechanism but precise comparisons on the production
at TEVATRON and LHC. Although some certain uncertainties will be
able to understand (control) better when the next to leading order
(NLO) pQCD calculation is achieved, considering that the NLO
calculation cannot be available soon due to its complicatedness,
we restrict ourselves here to examine the uncertainties only `up
to' the lowest order. The uncertainties that we will examine in
the paper, include the variations about $\alpha_s$-running, the
choices of the factorization energy scale, the various versions of
parton distribution functions (PDFs), the values of the bound
state parameters and {\it etc.}. To be useful references, the
production with possible kinematic cuts, which roughly `match' the
detectors at TEVATRON and at LHC, is also investigated.

It is known that at high energy hadronic colliders, numerous $B_c$
events may be produced, moreover due to the fact that the $B_c$
meson has a quite large branching ratio to decay into a $B_s$
meson (several tens percent)\cite{chen,dec,dec1}, thus, in
principle, copious and precisely tagged $B_s$ mesons at production
position may be collected through the inclusive decays of $B_c$ at
the hadronic colliders ($B_c$ ($\bar{B_c}$) is the charged meson,
as long as the charge of $B_c$ ($\bar{B_c}$) and/or its decay
products is determined, the produced $B_s$ ($\bar{B_s}$) mesons
through the decays of $B_c$ ($\bar{B_c}$) are tagged precisely at
the decay vertex of $B_c$ ($\bar{B_c}$)). To study $B_s-\bar{B_s}$
mixing and $CP$ violation are very interesting topics and can be
done only in hadronic environment. For experimental studies of
$B_s-\bar{B}_S$ mixing and certain $CP$ violation in $B_s$ decays,
the $B_s$ mesons being tagged at their production position are
crucial. In hadronic environment to obtain copious $B_s$ mesons
which are precisely tagged at their production positions, the way
through $B_c$ decays is worth for considering seriously especially
at LHC. In addition, potentially this way has a great advantage in
rejecting backgrounds and deducing the systematic errors
\cite{mord}. Therefore the feasibility for studying the $B_s$
mesons in this way should be investigated carefully, especially,
by means of Monte Carlo simulating the events under specific
detector environment so as to see the precise efficiency to have
the tagged $B_s (B_s)$ mesons and whether the idea is really
practicable or not.

The paper is organized: following Introduction, in Section II we
present the studies of the uncertainties in the estimates of the
hadronic production for the dominant mechanism of gluon-gluon
fusion and also a brief comparison between the mechanisms of
gluon-gluon fusion and quark-antiquark annihilation. In Section
III, we present the production with possible kinematic cuts, which
roughly match the detector situation at TEVATRON and LHC. We also
compute the production at TEVATRON with C.M. energy changes from
RUN-I to RUN-II. In the last section, we make some discussions and
a short summary.

\section{The Uncertainties in Estimates}

As stated in the introduction, there are two mechanisms for the
lowest order of pQCD $\alpha_s^4$ approach in the hadronic $B_c$
production, the gluon-gluon fusion and the quark-antiquark
annihilation. Of them the gluon-gluon fusion mechanism is the
dominant one, and there are 36 Feynman diagrams for its
subprocess. The typical Feynman diagrams for the subprocess are
plotted in FIG.1(a). In comparison, for the other mechanism, its
subprocess is of quark-antiquark annihilation and there are 7
Feynman diagrams. The typical ones are plotted in FIG.1(b). In
fact, if we consider the production only through the dominant
gluon-gluon fusion mechanism, then the contributions from the
quark-antiquark annihilation mechanism will be considered as one
kind of uncertainties. In this section, firstly we will focus on
the gluon-gluon fusion mechanism and then we will make a brief
comparison between these two mechanisms.

\begin{figure}
\setlength{\unitlength}{1mm}
\begin{picture}(80,90)(30,15)
\put(-10,-5){\includegraphics{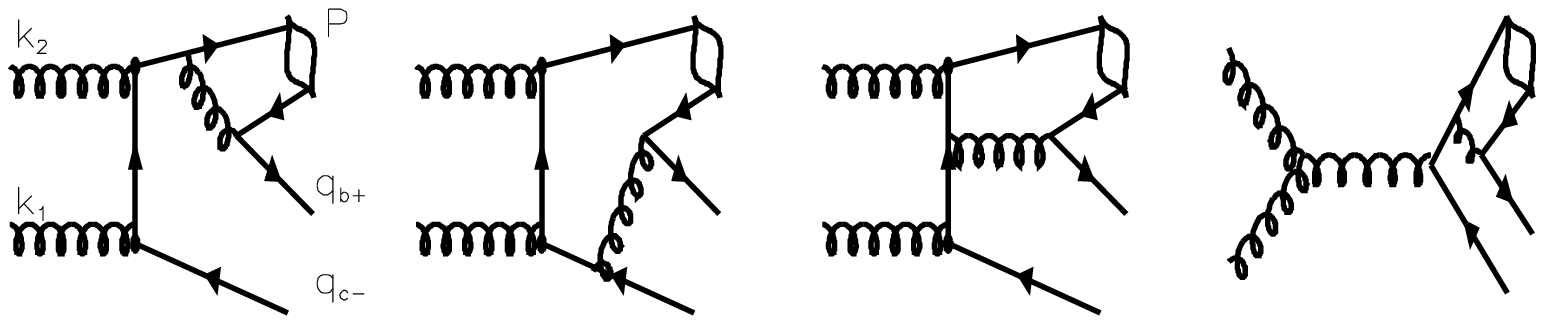}}
\put(-10,-25) {\includegraphics{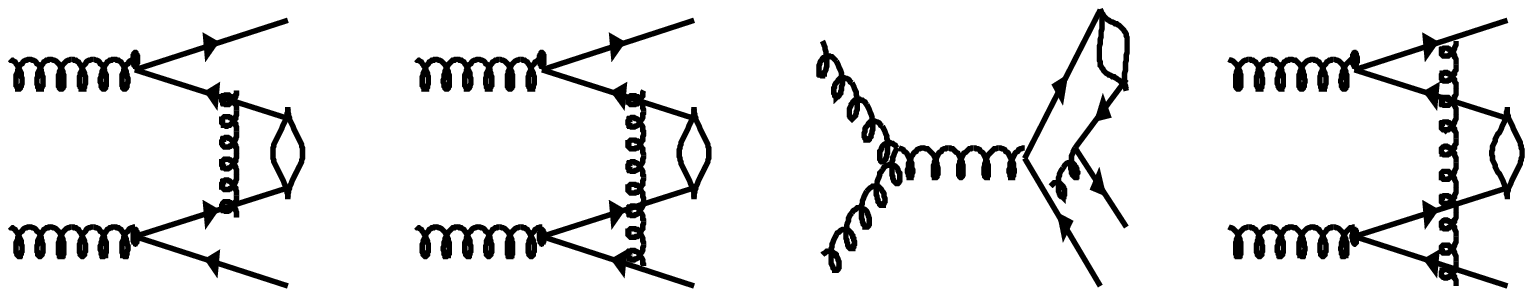}}
\put(-10,-45) {\includegraphics{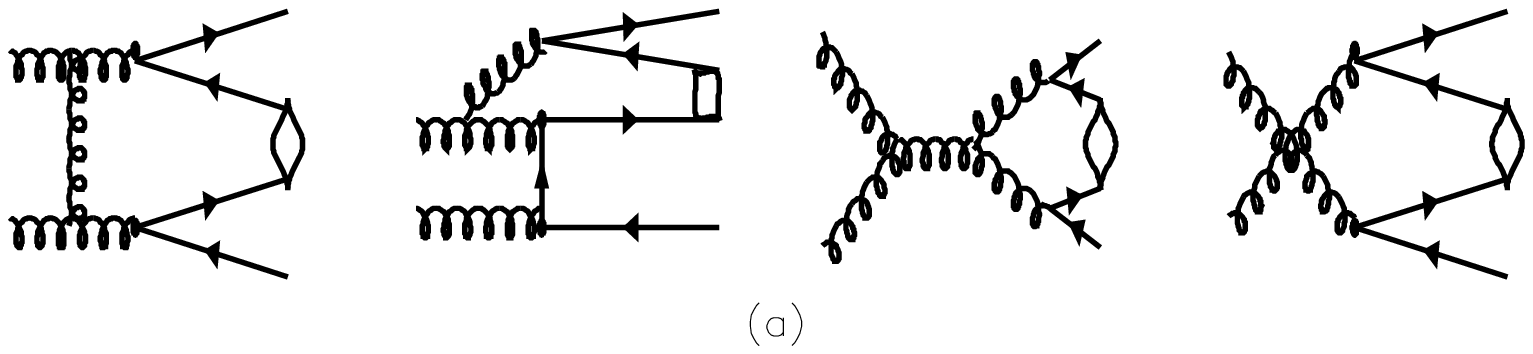}}
\put(-10,-75) {\includegraphics{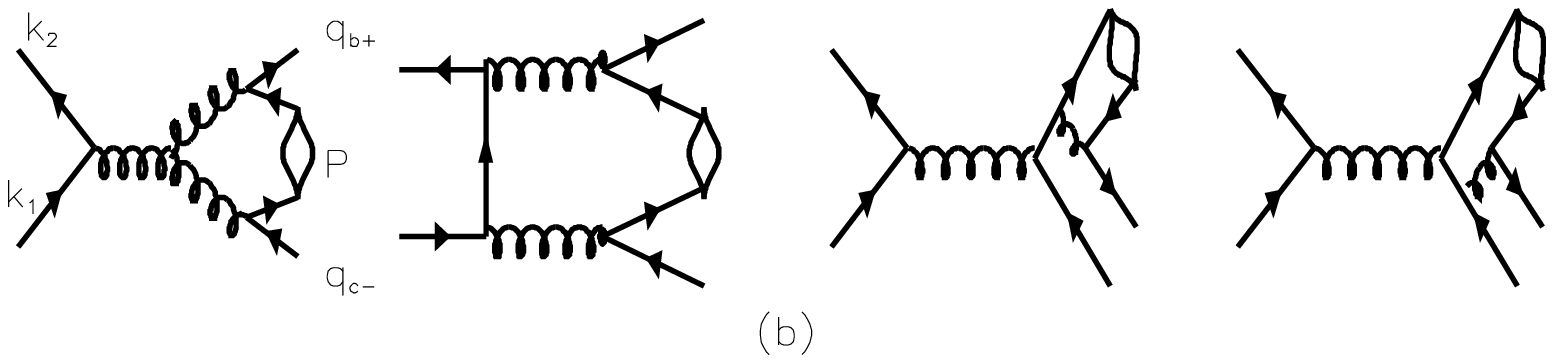}}
\end{picture}
\caption{(a). The typical Feynman diagrams for the gluon-gluon
fusion subprocess; (b). The typical Feynman diagrams for
quark-antiquark annihilation subprocess.} \label{feyn}
\end{figure}

To be more useful experimentally, we study the uncertainties of
$B_c$ hadronic production by means of the `complete approach',
i.e., at collision C.M. energy $\sqrt{S}$,
\begin{eqnarray}
d\sigma(S,p_T,\cdots)&=&\sum_{ij}\int dx_{1}\int
dx_{2}F^{i}_{H_1,P_{1}}(x_{1},\mu^2_{F})\cdot
F^{j}_{H_2,P_{2}}(x_{2},\mu^2_{F})\nonumber \\
&\cdot& d\hat{\sigma}_{ij\rightarrow
B_{c}b\bar{c}}(P_1,P_2,x_{1},x_{2},\mu^2_{F}, \hat{s},p_T,\cdots)\
, \label{cross}
\end{eqnarray}
where $F^{i}_{H_1,P_{1}}(x_{1},\mu^2_{F})$ and
$F^{j}_{H_2,P_{2}}(x_{2},\mu^2_{F})$ are the PDFs of incoming
hadrons $H_1$ (momentum $P_1$) and $H_2$ (momentum $P_2$) for
parton $i$ (with the momentum fraction $x_1$) and parton $j$ (with
the momentum fraction $x_2$) respectively; $\mu_F$ is the energy
scale where the factorization and renormalization for the PDFs and
the hard subprocess are made; $d\hat{\sigma}_{ij\rightarrow
B_{c}b\bar{c}}$ is the differential cross-section of the relevant
hard subprocess, in which $\hat{s}=x_1x_2S$ is the c.m.s. energy
of the subprocess and $p_T$ is the transverse momentum of $B_c$.
In the sub-sections below, we discuss the uncertainties which are
from the non-perturbative factors and from those related to the
hard subprocess.

\subsection{The Uncertainties Relevant to the Parameters of Potential Model
and Masses of Quarks and the Meson}

For all the approaches and the color singlet mechanisms to the
$B_c$ production, the `decay constant' $f_{B_{c}}$ relating the
hadron matrix element, as one of important input-parameters, is
needed as input, and so far it may be calculated either by
potential model \cite{spec} or by lattice QCD\cite{latt}, while in
the parer, we will make discussions related to it on potential
model only. Within potential model, it directly relates to the
wave function at the origin of the binding system in terms of the
following formula,
\begin{equation}
f^{2}_{B_{c}}=\frac{12|\Psi_{[1S]}(0)|^{2}}{m_{B_c}}\ ,
\end{equation}
where $m_{B_c}$ is the $B_c$ meson mass, $\Psi_{[1S]}$ is the wave
function at the origin of the binding system $(c\bar{b})$ at the
$[1S]$ level. Since the spin splitting effects are ignored here,
so there is no difference for the decay constant between the spin
stats $[^1S_0]$ and $[^3S_1]$. There are several parameters in the
potential model, which need to be fixed by fitting the
experimental data of the heavy quarkonia ($(c\bar{c})$ and
$(b\bar{b})$), so when computing the spectrum and the wave
functions of the $(c\bar{b})$ system\cite{spec}, certain
uncertainties from the fitting procedure will cause some
uncertainties of $f_{B_{c}}$. In the leading order (LO)
approximation for the production, $f_{B_{c}}$ appears in the
amplitude as a linear factor exactly, so the production cross
sections are proportional to it squared. Therefore, the
uncertainties in the production from $f_{B_c}$ can be figured out
straightforwardly, so throughout the paper, we will fix the value
$f_{B_c}=0.48$ GeV for the ground states (the relevant
uncertainties would be figured out if need).

In addition to the decay constant $f_{B_c}$, the quark mass values
$m_c$ and $m_b$ also `generate' uncertainties for the hadronic
production. If all of the rest parameters in the potential might
be fixed, then the values of quark masses would determine the
decay constant $f_{B_c}$, the meson mass $m_{B_c}$ and {\it etc.}
completely. However at present, the parameters in potential model,
which also appear in the formula Eq.(\ref{cross}), cannot be
completely fixed by fitting the available data of the heavy
quarkonia, so the relations of the quark masses to the decay
constant $f_{B_c}$, the meson mass and {\it etc.} cannot be well
determined. Therefore, when studying the uncertainties for the
hadronic production, we can consider all of the factors,
explicitly appearing in the potential model and in the formula for
the subprocess Eq.(\ref{cross}), in a `factorized' way.
Furthermore, since $B_c$ is the non-relativistic and weak-binding
bound state, at LO the relative momentum between the constitute
quarks can be ignored, e.g., we approximately have $m_{B_c}\simeq
m_b+m_c$.

\begin{table}
\begin{center}
\caption{The total cross section (in unit nb) for hadronic
production of $B_c[1^{1}S_{0}]$ ($B_c^*[1^{3}S_{1}]$) with various
values of the $c$-quark mass $m_c$ and fixed $b$-quark mass
$m_b=4.9$ GeV. For definiteness and focussing the uncertainties
from the quark masses alone, the rest uncertainty sources are
fixed: the mass of the meson $B_c$, $m_{B_c}=m_c+m_b$; the gluon
distribution function is taken from CTEQ5L; the factorization
energy scale is chosen $\mu_F^2=Q^{2}=\hat{s}/4$ and the running
$\alpha_s$ is of leading order.} \vskip 0.6cm
\begin{tabular}{|c||c|c|c|c|c|c||c|c|c|c|c|c|}
\hline - & \multicolumn{6}{|c||}{~~~TEVATRON~($\sqrt S=2.$
TeV)~~~}& \multicolumn{6}{|c|}{~~~LHC~($\sqrt S=14.$ TeV)~~~}\\
\hline\hline $m_c$ (GeV) & ~~$1.3$~~ & ~~$1.4$~~ & ~~$1.5$~~
&~~$1.6$~~&~~$1.7$~~ & ~~$1.8$~~ & ~~$1.3$~~ & ~~$1.4$~~
 & ~~$1.5$~~ &~~$1.6$~~ &~~$1.7$~~ &~~$1.8$~~\\
\hline $\sigma_{B_{c}}(nb)$ & 4.84 & 3.87 & 3.12 &
2.56 & 2.12 & 1.76 & 75.6 & 61.0 & 49.8 & 41.4 & 34.7 & 28.9 \\
\hline $\sigma_{B^{*}_c}(nb)$ & 12.3 & 9.53 & 7.39 & 5.92 &
4.77 & 3.87 & 194. & 153. & 121.& 97.5 & 80.0 & 66.2 \\
\hline\hline
\end{tabular}
\label{tabmc}
\end{center}
\end{table}

Throughout the paper, we study the uncertainties in `a
factorization way', i.e., all of the parameters vary independently
in their reasonable regions. For instance, when focussing on the
uncertainties from $m_c$, we let it be a basic `input' parameter
varying in a possible range
\begin{equation}
1.3 GeV \leq m_{c} \leq 1.8 GeV, \label{c-mass}
\end{equation}
with all the other factors, including the $b$-quark mass, the
decay constant  $f_{B_c}$ and {\it etc.} being fixed.

The uncertainties from $m_c$ are indicated by the calculated total
cross sections with the $m_c$ of Eq.(\ref{c-mass}), which are
given in TABLE~\ref{tabmc}. Note that for the mass of $B_c$, the
experimental result is $m_{B_c}=6.4\pm 0.4$ GeV \cite{CDF1}, while
the prediction by the potential model gives $m_{B_c}=6.1 \sim 6.3$
GeV \cite{spec} and that by lattice QCD is about $6.4$ GeV
\cite{latt}. Thus with $m_b=4.9$ GeV and $m_{B_c}\simeq m_b+m_c$,
the obtained $m_{B_c}$ is in the region of theoretical prediction
as well as experimental measurement. The uncertainties from $m_b$
can be analyzed in a similar way, where the $b$-quark mass runs
over the range:
\begin{equation}
4.5 GeV \leq m_{b} \leq 5.3 GeV. \label{b-mass}
\end{equation}
The results for various b-quark mass are put in TABLE~\ref{tabmb}.
In Tables~\ref{tabmc} and \ref{tabmb}, the total cross-section for
the hadronic production of $B_c[1^{1}S_{0}]$ and
$B_c^*[1^{3}S_{1}]$ at TEVATRON and LHC are computed, where the
other factors are fixed precisely as: the PDFs are taking as
CTEQ5L \cite{5lcteq}; the strong coupling $\alpha_s$ is in leading
order and the factorization energy scale is taken to be
$\mu_F^2=Q^{2}=\hat{s}/4$, where $\hat{s}$ is the C.M. energy
squared of the subprocess.

\begin{table}
\begin{center}
\caption{The total cross section (in unit nb) for hadronic
production of $B_c[1^{1}S_{0}]$ and $B_c^*[1^{3}S_{1}]$ with
various values of the $b$-quark mass $m_b$ and fixed $c$-quark
mass $m_c=1.5$ GeV. For definiteness and focussing the
uncertainties from the quark masses alone, the gluon distribution
function is taken from CTEQ5L; the factorization energy scale is
chosen $\mu_F^2=Q^{2}=\hat{s}/4$ and the running $\alpha_s$ is of
leading order.} \vskip 0.6cm
\begin{tabular}{|c||c|c|c|c|c||c|c|c|c|c|}
\hline\hline - & \multicolumn{5}{|c||}{~~~TEVATRON~($\sqrt S=2.$
TeV)~~~}& \multicolumn{5}{|c|}{~~~LHC~($\sqrt S=14.$ TeV)~~~}\\
\hline\hline $m_b$(GeV) & ~~~$4.5$~~~ & ~~~$4.7$~~~ & ~~~$4.9$~~~
&~~~$5.1$~~~&~~~$5.3$~~~& ~~~$4.5$~~~ & ~~~$4.7$~~~
& ~~~$4.9$~~~ & ~~~$5.1$~~~ & ~~~$5.3$~~~\\
\hline $\sigma_{B_{c}}(nb)$ & 4.10 & 3.55 & 3.12 & 2.70 & 2.38&
63.4 & 56.2 & 49.8 & 44.1& 39.6\\
\hline $\sigma_{B^{*}_c}(nb)$ & 9.43 & 8.35 & 7.39 & 6.59 & 5.89&
148. & 133. & 121.& 110. & 100. \\
\hline\hline
\end{tabular}
\label{tabmb}
\end{center}
\end{table}

From the tables, one may observe how the values of $m_c$ and $m_b$
affect the cross section up to a sizable degree. Roughly speaking,
both at TEVATRON and LHC, when $m_b$ increases by steps of $0.2$
GeV, then the cross section decreases by $10\%\sim 20\%$, while
when $m_c$ increases by steps of $0.1$ GeV, the cross section also
decreases by $10\%\sim 20\%$.

In further investigations of the paper when examining the
uncertainties from the rest factors, we shall take the values,
$m_{c}=1.5$ GeV and $m_b=4.9$ GeV, for the quark masses.

\subsection{Uncertainties Relevant to the QCD Parameters and
PDFs}

In this sub-section, let us focus on the gluon-gluon fusion
mechanism only.

As is shown in Eq.(\ref{cross}), PDFs
$F^{i}_{H_1(P_{1})}(x_{1},\mu^2_{F})$ and
$F^{j}_{H_2(P_{2})}(x_{2},\mu^2_{F})$ with $H_1$=proton,
$H_2$=proton for LHC; $H_1$=proton, $H_2$=anti-proton for TEVATRON
generate uncertainties in the production. PDFs are of
non-perturbative nature, and in Eq.(\ref{cross}) they have been
factorized out at the energy scale $\mu_F^2$ with the help of pQCD
factorization theorem. Usually the factorization is carried out at
$Q^2$, a characteristic energy scale for the hard subprocess, i.e.
$\mu_F^2=Q^2$, so we use $Q^2$ and $\mu_F^2$ on the equal footing
throughout the paper.

As stated in the Introduction, all the estimates of the production
are of leading (the lowest) order of pQCD and then how to choose
the energy scale $Q^2$ is a very tricky problem. If $Q^2$ is
chosen properly, the results may be quite accurate\footnote{In
PYTHIA the value of $Q^2$ can be definitely fixed at each
interaction vertex, therefore when applying PYTHIA to the
production, there is no energy scale ambiguity. However, in PYTHIA
all kinds of $b$-hadron ($B_s, B_c, \Lambda_b, \cdots$) events
will be produced according to the proper fragmentation
possibilities, among which the fragmentation possibility for $B_c$
is quite small \cite{prod0,prod}, thus in terms of PYTHIA to
generate the $B_c$ events is not `economical', i.e., when
simulating the $B_c$ events, enormous unwanted $b$-hadron events
with higher probability will be generated in the meantime. By
contrast, the approach which we are adopted here has quite high
efficiency to generate the $B_c$ events, though the uncertainties
from the choices of $Q^2$ are unavoidable.}. From experience, for
a hard subprocess with a two-body final state, generally when
taking $Q^2=\frac{1}{4}\hat{s}$ ( $\sqrt{\hat{s}}$ is the C.M.
energy of the subprocess), we can achieve quite an accurate result
for LO calculations. However, in the present case, the gluon-gluon
fusion subprocess is of three-body final state and contain heavy
quarks, so there are ambiguities in choosing $Q^2$ and various
choices of $Q^2$ would generate quite different results. Since
such kind of ambiguity cannot be justified by the LO calculation
itself, so we take it as the uncertainty of the LO calculation,
although when the NLO calculation of the subprocess is available,
the uncertainty will become under control a lot. While the NLO
calculation is very complicated and it cannot be available in the
foreseeable future, so here we take $Q^2$ as a possible
characteristic momentum of the hard subprocess being squared.
Having reviewed the choices of $Q^2$ in the literature, in the
following we choose a few typical examples to study this kind of
uncertainties.

According to the factorization, the running of $\alpha_s$ and PDFs
should be of leading logarithm order (LLO), and the energy scale
$Q^2$ appearing in the calculation should be taken as one of the
possible characteristic energy scales of the hard subprocess. In
principle, as has been shown in references
\cite{prod0,prod,prod1,prod2,prod3,avaa,prod4}, there are several
typical ways to choose the energy scale $Q^2$ and to relate it to
a characteristic energy scale ($Q^2\gg \Lambda_{QCD}^2$) of the
subprocess, such as $Q^2=4m^2_c$, $Q^2=4m_b^2$,
$Q^2=p_T^2+m_{B_c}^2$ (the `transverse mass' squared of $B_c$),
$Q^2=\frac{1}{4}\hat{s}$ and {\it etc.}. When studying the
uncertainties on the various choices of $Q^2$, we precisely choose
the following four types of $Q^2$:

Type A: $Q^2=\hat{s}/4\,,$ the C.M. energy squared of the
subprocess that is divided by $4$;

Type B: $Q^2=p_{T}^2+m_{B_c}^2\,,$ the transverse mass squared of
the $B_c$ meson;

Type C: $Q^2=\hat{s}\,,$ the C.M. energy squared of the
subprocess;

Type D, $Q^2=p_{Tb}^2+m_b^2\,,$ the transverse mass squared of the
$b$ quark.

Since the PDFs can be obtained only through global fitting of the
experimental data and evolute them to the requested characteristic
scale $Q^2$ in standard way of pQCD, so there are several groups,
CTEQ \cite{5lcteq,6lcteq}, GRV \cite{98lgrv} and MRS
\cite{2001lmrst} etc, who devote themselves to offer accurate PDFs
to the world, and to keep PDFs updated as soon as possible when
new relevant experimental data are available. Thus in literature,
different versions of PDFs (including different issues by the same
group) are used in the estimates of the hadronic production. To
study the uncertainties due to various versions of PDFs, we
consider the above mentioned three groups.

The generator BCVEGPY is programmed based on the dominant
subprocess to LO pQCD only, to be self-consistent, when applying
BCVEGPY we shall adopt the LO PDF. Thus we take CTEQ6L
\cite{6lcteq}, CTEQ5L \cite{5lcteq}, GRV98L \cite{98lgrv} and
MRST2001L\cite{2001lmrst} as typical examples for PDFs. The
versions of the gluon distributions ended with `L' are accurate
up-to the leading logarithm order (LLO), i.e., their QCD evolution
effects with $\alpha_s$ running are included, so for the
production to show the uncertainties correctly up to LO accuracy,
it is necessary for the PDFs, the hard subprocess and the QCD
`coupling constant' $\alpha_s$ `run' to the energy scale $Q^2$
properly. The running $\alpha_s$ in the PDFs and in the subprocess
should be the same, and then when computing the production and
taking the PDFs from one version of the three groups, the running
$\alpha_s$ should also be taken from the same group.

For comparison between TEVATRON and LHC and to pinpoint the
uncertainties from PDFs, $\alpha_s$ running and the choices of the
characteristic energy scale $Q^2$, we calculate the production
cross sections according to four types of PDFs, the strong
coupling $\alpha_s$ up to LLO fixed by the PDFs and the
characteristic $Q^2$ chosen as Type A and Type B. The obtained
results are shown in TABLE~\ref{tab321}. From the table, we can
see that the gluon distribution functions, the running coupling
constant and the corresponding energy scale, all affect the cross
section to certain degrees.

\begin{table}
\begin{center}
\caption {Total cross-section for the hadronic production of
$B_c[1^{1}S_{0}]$ and $B_c^*[1^{3}S_{1}]$ at TEVATRON and at LHC
with the leading order (LLO) running $\alpha_s$ and the
characteristic energy scale $Q^2=\hat{s}/4$ or
$Q^2=p_{T}^2+m_{B_c}^2$. The cross section is in unit of nb.}
\vskip 0.1cm \arraycolsep15in
\begin{tabular}{|c||c|c|c|c||c|c|c|c|}
\hline\hline
 & CTEQ5L & CTEQ6L & GRV98L & MRST2001L& CTEQ5L &
 CTEQ6L & GRV98L & MRST2001L \\
\hline - &\multicolumn{4}{|c||}{$Q^2=\hat{s}/4$} &
\multicolumn{4}{|c|}{$Q^2=p_{T}^2+m_{B_c}^2$}\\
\hline\hline - & \multicolumn{8}{|c|}{TEVATRON} \\\hline
$\sigma_{B_c(1^{1}S_{0})}$& 3.12& 3.79 &
3.27 & 3.40 & 4.39 & 5.50  & 4.54 & 4.86\\
\hline $\sigma_{B^*_c(1^{3}S_{1})}$ & 7.39 & 9.07 &
7.88 & 8.16 & 10.7 & 13.4 & 11.1 & 11.9\\
\hline \hline - & \multicolumn{8}{|c|}{LHC}\\\hline
$\sigma_{B_c(1^{1}S_{0})}$ & 49.8 & 53.1 &
53.9 & 47.5 & 65.3 & 71.1 & 70.0 & 61.4 \\
\hline $\sigma_{B^*_c(1^{3}S_{1})}$ & 121.& 130. &
131.& 116. & 164. & 177.& 172. & 153.\\
\hline \hline
\end{tabular}
\label{tab321}
\end{center}
\end{table}

For all of the LO PDFs, the differences in the corresponding cross
sections given by them are small (about $10\%-20\%$). But the
choice of $Q^2$ as Type A and Type B causes the cross section to
change by a factor of about $\frac{1}{4}$ to $\frac{1}{3}$, which
is a comparatively large effect. From TABLE~\ref{tab321}, one may
see that the cross section at TEVATRON and at LHC presents a
somewhat different aspect of the cross section. Roughly speaking
at TEVATRON CTEQ6L is the biggest, while at LHC, GRV98L is the
biggest, under a certain characteristic energy scale such as $Q^2$
in type A.

From TABLE~\ref{tab321}, we may also see that the cross section of
the $B_c$ meson production at LHC are at least one order larger in
magnitude than that at TEVATRON. This is mainly due to the fact
that LHC has much higher collide energy than TEVATRON, so the
lowest boundary of the momentum fractions $x_i \;\; (i=1,2)$ at
LHC are much smaller than that at TEVATRON and then there are more
interacting partons i.e. gluons, which have a C.M. energy above
the threshold for the subprocess, in the collision hadrons at LHC
than at TEVATRON.

When the produced $B_c$ mesons with a small $p_T$ are too close to
the collision beam, they cannot be measured experimentally, so we
calculate the cross-sections with transverse-momentum cuts
$p_{Tcut}$. For comparison, we define the following ratio of the
integrated cross sections for TEVATRON and LHC,
$R=\left(\frac{\sigma_{TEVATRON}}{\sigma_{LHC}}\right)_{p_{Tcut}}$
and put the results for $B_c[1^{1}S_{0}]$ in TABLE~\ref{tab-jia}.

\begin{table}
\begin{center}
\caption{Dependence of the ratio
$R=\left(\frac{\sigma_{TEVTRON}}{\sigma_{LHC}}\right)$ on
$p_{Tcut}$ for $B_c[1^{1}S_{0}]$, where $Q^{2}=\hat{s}/4$, the
running $\alpha_s$ is in LO and the gluon distribution is CTEQ5L.}
\vskip 0.6cm
\begin{tabular}{|c||c|c|c|c||c|c|c|c|c|}
\hline\hline
- & \multicolumn{4}{|c||}{$B_c$} & \multicolumn{4}{|c|}{$B_c^*$}\\
\hline
$P_{Tcut}(GeV)$ & ~~~0~~~ & ~~~5~~~ & ~~~50~~~ & ~~~100`~~
& ~~~0~~~ & ~~~5~~~ & ~~~50~~~ & ~~~100~~~\\
\hline\hline $R (\times 10^{-2})$
& 6.26 & 5.24 & 1.00 & 0.31 & 6.09 & 5.19 & 0.97 & 0.24 \\
\hline\hline
\end{tabular}
\label{tab-jia}
\end{center}
\end{table}

To show the uncertainties due to various factors more
quantitatively, let us compute the differential cross sections of
$p_T$ and $y$ for the pseudo-scalar $B_c$ meson and draw the
curves accordingly.

\begin{figure}
\centering
\hfill\includegraphics[width=0.43\textwidth]{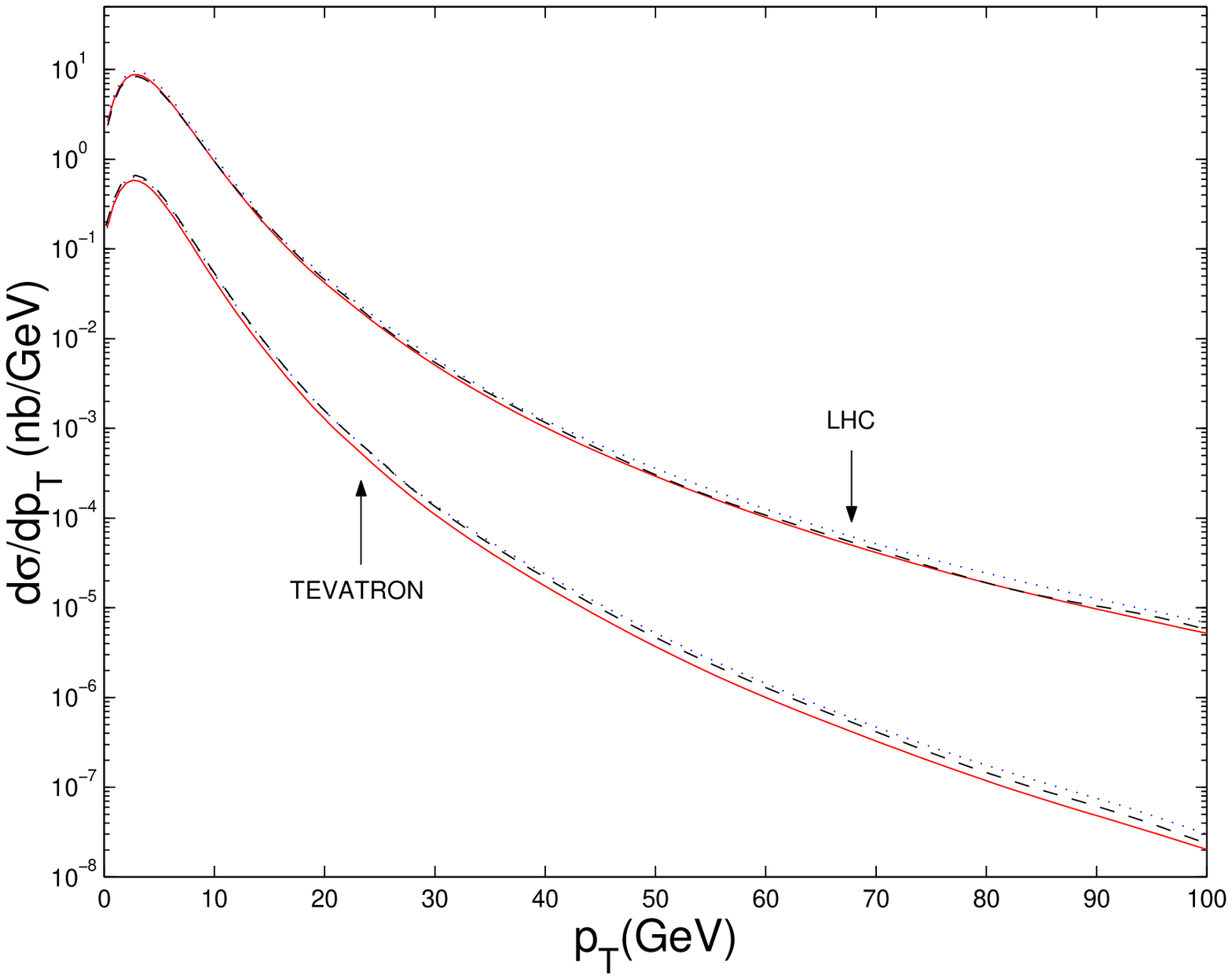}%
\includegraphics[width=0.43\textwidth]{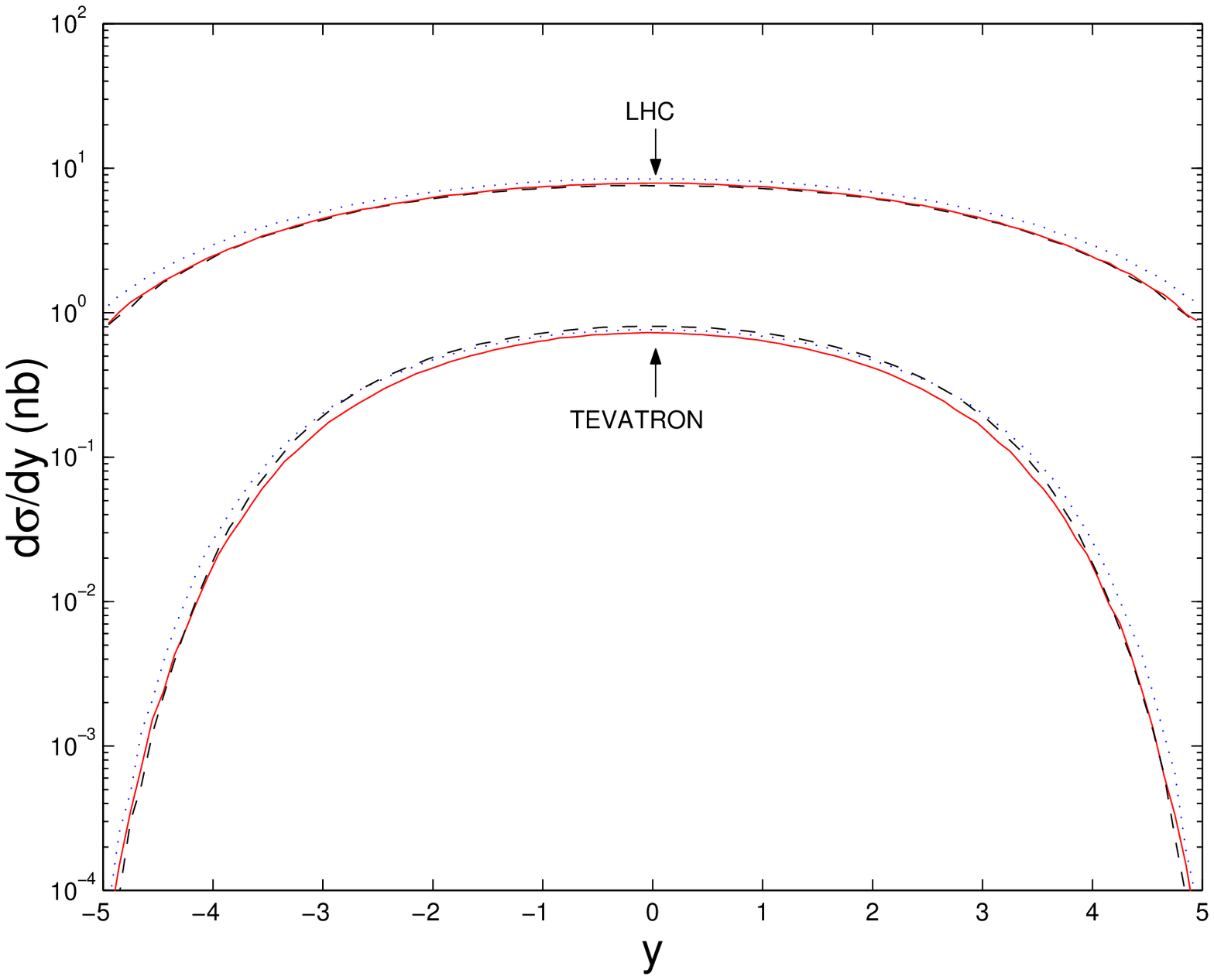}\hspace*{\fill}
\caption{$B_c$ differential distributions versus its transverse
momentum $p_T$ and rapidity $y$ for different versions of gluon
distributions in leading order. The characteristic energy scale is
taken as $Q^2=\hat{s}/4$. Solid line: CTEQ5L; dotted line: GRV98L
and dashed line: MRST2001L. The upper (lower) three lines
corresponding to the distributions in LHC (TEVATRON).}
\label{pdft}
\end{figure}

\begin{figure}
\centering
\hfill\includegraphics[width=0.43\textwidth]{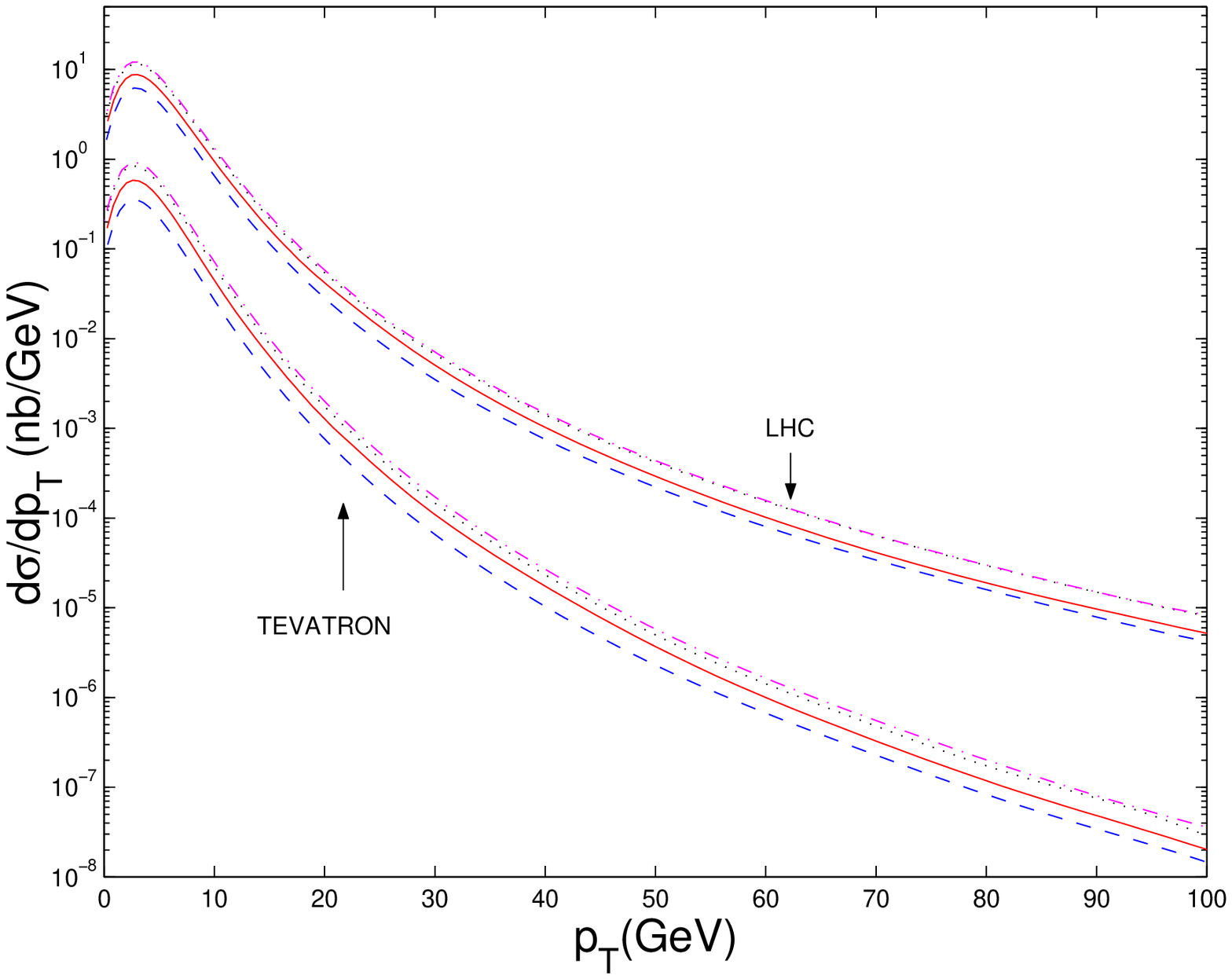}%
\includegraphics[width=0.43\textwidth]{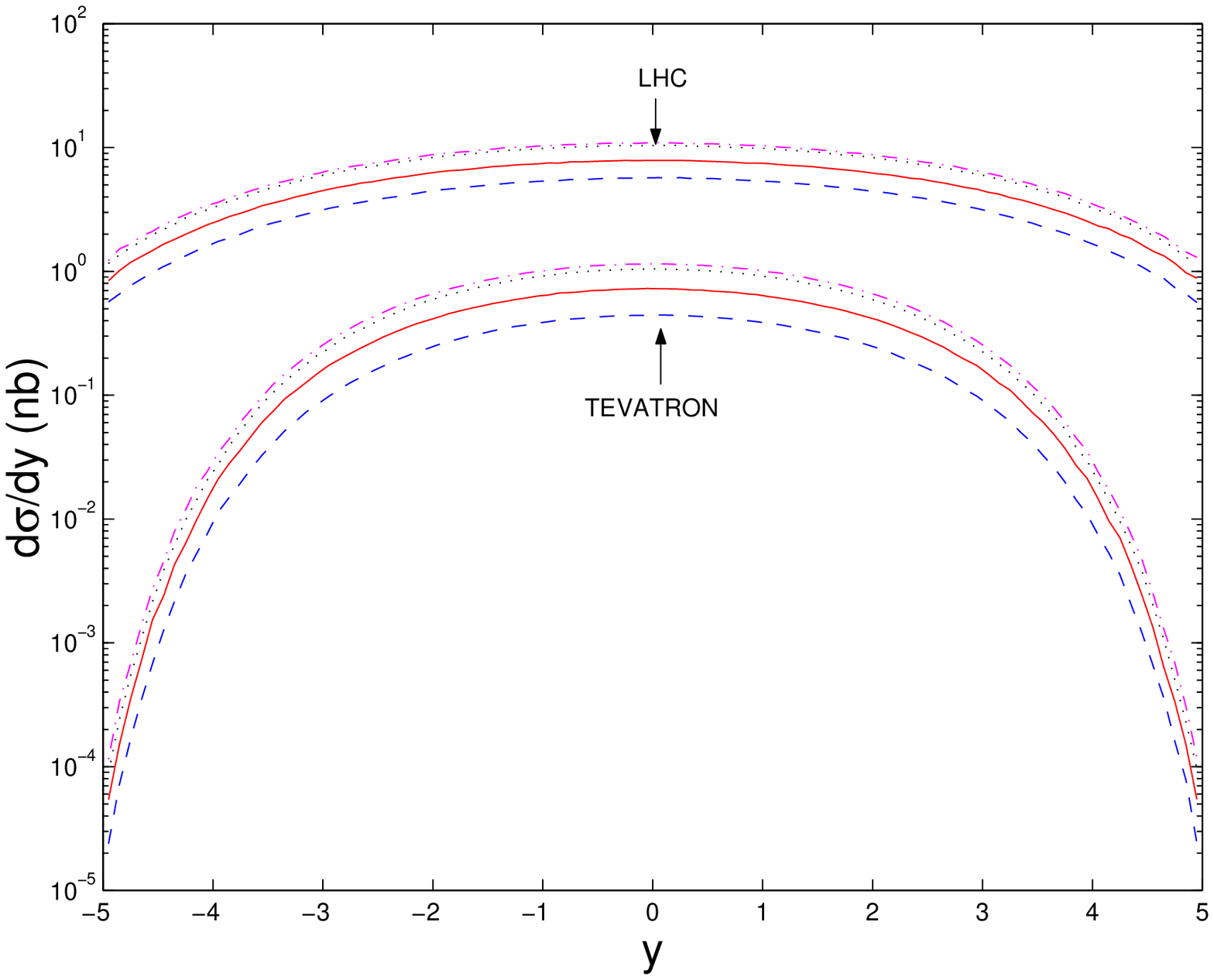}\hspace*{\fill}
\caption{$B_c$ differential distributions versus its transverse
momentum $p_T$ and rapidity $y$ for four typical choices of the
characteristic energy scale $Q^2$. The gluon distribution is
chosen as CTEQ5L and the running $\alpha_s$ is in leading order.
The choice of $Q^2$ is: solid line Type A; dotted line Type B;
dashed line Type C and dash-dot line Type D. The upper (lower)
four lines corresponding to the distributions at LHC (TEVATRON).}
\label{cteqQtpt}
\end{figure}

\begin{figure}
\centering
\hfill\includegraphics[width=0.43\textwidth]{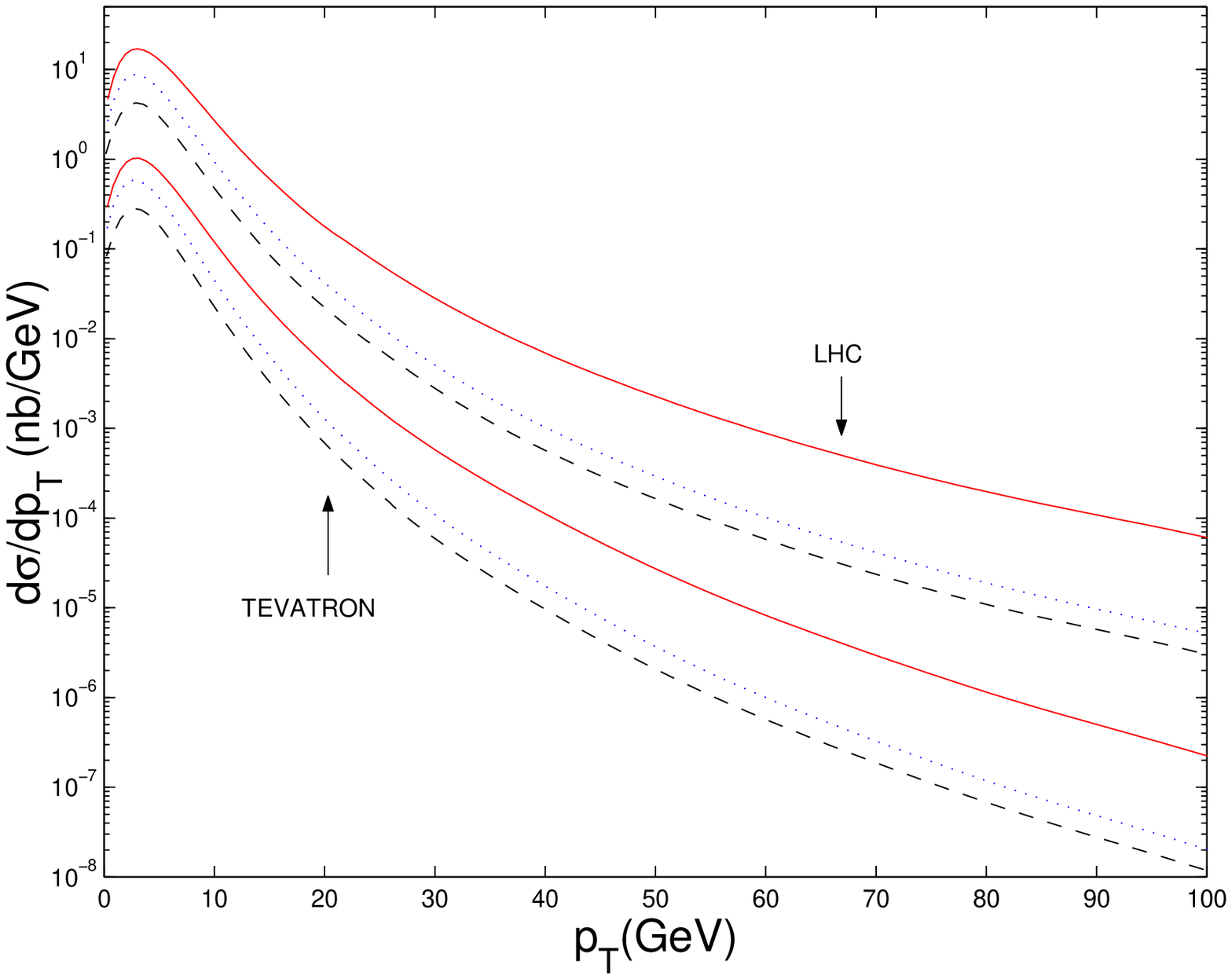}%
\includegraphics[width=0.43\textwidth]{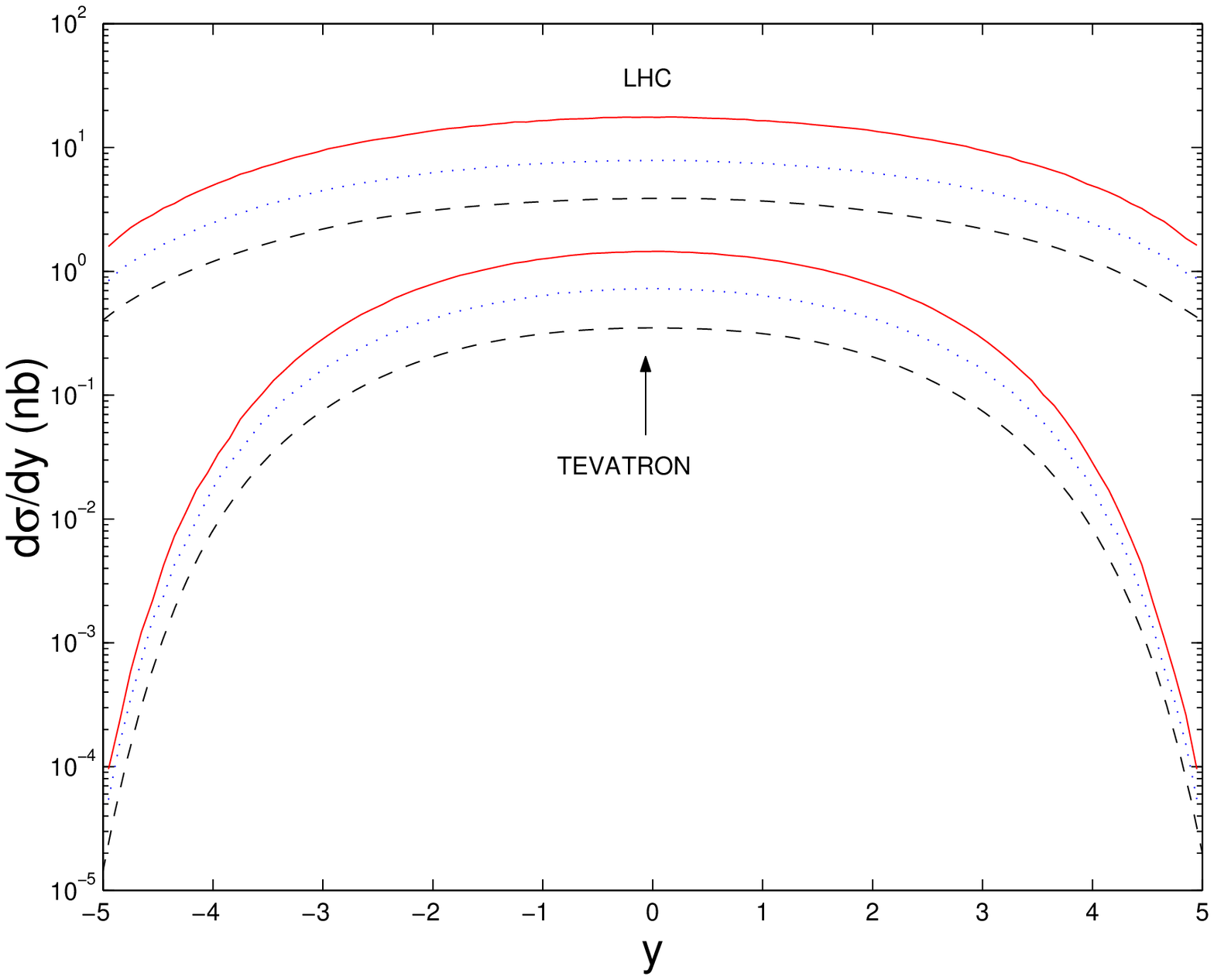}\hspace*{\fill}
\caption{$B_c$  differential distributions versus its transverse
momentum $p_T$ and rapidity $y$ for different running $\alpha_s$.
The gluon distribution of CTEQ5L and the Type A characteristic
energy scale $Q^2=\hat{s}/4$, are used here. The solid line stands
for the constant $\alpha_s=0.22$, the dotted line for the leading
order running $\alpha_s$ and the dashed line for the next to
leading order running $\alpha_s$. The upper (lower) four lines
correspond to the distributions in LHC (TEVATRON).}
\label{cteqStpt}
\end{figure}

From FIG.~\ref{pdft}, we may see that the differential
distributions for the three PDFs CTEQ5L, GRV98L and MRST2001L are
very similar. The corresponding differences for TEVATRON and LHC
are evident (the differences as shown in TABLE~\ref{tab321} are
less than $20\%$). In regions of comparatively small $p_T$ and $y$
in TEVATRON the distributions of $p_T$ and $y$ obtained with
MSRT2001L are the largest, and then those of GRV98L and CTEQ5L;
while at LHC, those of GRV98L are the largest, and then those of
CTEQ5L and MSRT2001L. From the figure, we also see that the $p_T$
distributions in TEVATRON are steeper than those in LHC.

Since in literature, various $\alpha_s$ have been adopted in
making estimates of the $B_c$ production
\cite{prod,prod1,prod2,prod3,avaa,prod4} so to follow these
authors, we try several $\alpha_s$ to compute the production.
Namely, in addition to the LO $\alpha_s$, the constant
$\alpha_s=0.22$ and the NLO $\alpha_s$ are also taken into the
computations and the curves of the production obtained by the
various $\alpha_s$ are drawn in FIG.~\ref{cteqStpt}.

We may see from FIG.~\ref{cteqStpt} that the variations of the
results are quite large for both TEVATRON and LHC; those with
$\alpha_s=0.22$ are the largest, those with NLO $\alpha_s$ are the
smallest, while those with LO $\alpha_s$ are in the middle.

In summary, of all the uncertainties which have been studied in
this subsection, the uncertainty caused by the different choice of
$Q^2$ is the largest, that can be as great as a factor around
$\frac{1}{3}$ (although in the literature results with various
choices of $\alpha_s$ running have given rise to even a greater
change).

\subsection{Comparison between the Mechanisms of Gluon-Gluon Fusion and
Quark-Antiquark Annihilation}

In this subsection we make a brief comparison between the
mechanisms of gluon-gluon fusion and quark-antiquark annihilation.
Since, in a hadron such as $P, \bar{P}, \pi, \cdots$, the PDFs of
heavy quarks are very small throughout the active region
$|x_i|\leq 1$, it is sufficient to take into account only the
light quark pair ($u$-$\bar{u}$, $d$-$\bar{d}$ and $s$-$\bar{s}$)
annihilations.

\begin{figure}
\centering
\includegraphics[width=0.43\textwidth]{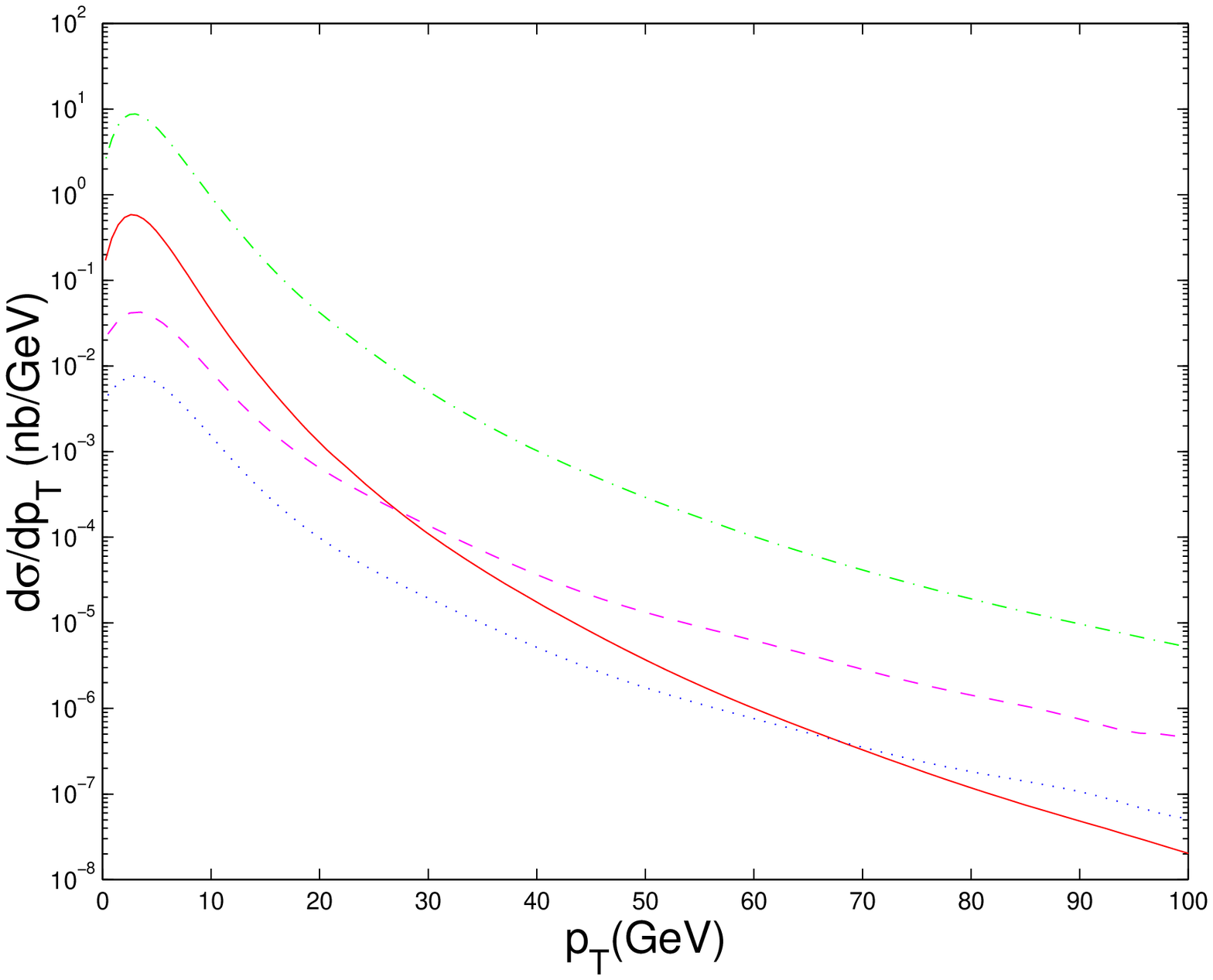}
\includegraphics[width=0.43\textwidth]{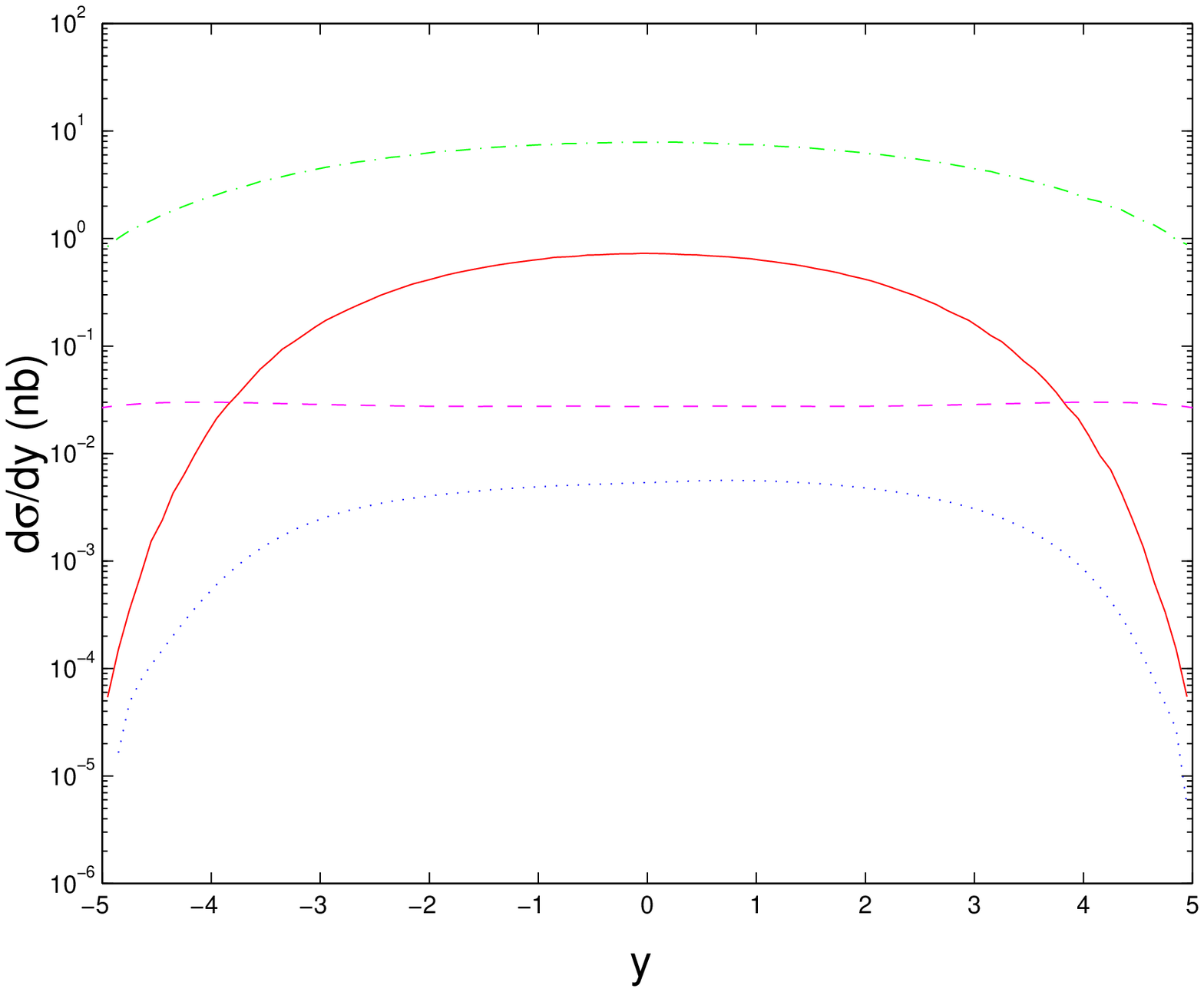}
\caption{$B_c$ distributions versus its transverse-momentum $p_T$
and rapidity $y$ of $B_c$ due to gluon-gluon fusion and
quark-antiquark annihilation (the direction of positive $y$ is
defined as that of the momentum of the colliding proton). The PDFs
are taken from CTEQ5L, $\alpha_s$ running is set at leading order,
and $Q^2=\frac{1}{4}\hat{s}$. The solid lines correspond to
gluon-gluon fusion and the dotted lines to quark-antiquark
annihilation in TEVATRON. The dashed-dotted lines correspond to
gluon-gluon fusion and the dashed lines to quark-antiquark
annihilation in LHC.} \label{ggqq}
\end{figure}

The comparative results for the gluon-gluon fusion and the
quark-antiquark annihilation mechanisms.In FIG\,.\ref{ggqq}, we
show

From FIG\,.\ref{ggqq} we see that for production at a high energy,
such as TEVATRON and LHC, the gluon-gluon fusion mechanism is
dominant over the quark-antiquark annihilation mechanism,
typically when the transverse momentum $p_T \leq 60$ GeV and
rapidity $|y|\leq 5.0$. Only in TEVATRON do the contributions from
gluon-gluon fusion become smaller than those from quark-antiquark
annihilation when $p_T\geq 70$ GeV. Generally speaking, for the
transverse momentum distributions, at low $p_T$ the contributions
from gluon-gluon fusion are greater than those from
quark-antiquark annihilation by at least two orders of magnitude
for both LHC and TEVATRON. For high transverse momentum ($p_T\geq
70$ GeV), the difference in the two mechanisms changes sign in
TEVATRON (i.e. contributions from quark-antiquark annihilation
become greater than those from gluon-gluon fusion due to the fact
that with such a high $p_T$, the valence quark PDFs play a
significant role), but in LHC the contributions from gluon-gluon
fusion are still greater by an order of magnitude than those from
quark-antiquark annihilation. For rapidity $y$ distributions, in
the $|y|\simeq 0$ region the contributions from gluon-gluon fusion
are greater by two orders of magnitude in both colliders; but in
TEVATRON in the higher region of $2.0\leq |y|\leq 5.0$,
contributions from the valance quarks' annihilations become more
important, so that contributions from quark-antiquark annihilation
increase remarkably (though still below those from gluon-gluon
fusion). It is interesting to note that there is a slight
asymmetry in the rapidity $y$ of the meson $B_c$ for the
quark-antiquark annihilation subprocess. In TEVATRON it is of $P
\bar{P}$ collision, so the asymmetry in $y$ is maintained somewhat
(see FIG.\ref{ggqq}). We will describe the feature in TEVATRON
elsewhere\cite{changwu}.

The contributions from quark-antiquark annihilation mechanism to
$B_c$ hadronic production, compared with those from gluon-gluon
fusion mechanism, can be ignored safely in LHC in the allowed
kinematic region. In TEVATRON, they can also be ignored over most
of the allowed kinematic region, i.e., only a little allowed
kinematic region is an exception. In the paper, therefore, we
consider the gluon-gluon fusion mechanism mainly.

\section{Kinematic cuts for the hadronic production of $B_c$ meson }

Experimentally, there is no detector which can cover all the
kinematics of the events, so only some of $B_c$ produced events
can be observed completely. For instance, in a high energy
hadronic collider, the $B_c$ events with a small $p_T$ and/or a
large rapidity $y$ (the produced $B_c$ mesons move very close to
the beam direction) cannot be detected by the detectors at it
directly, so this kind of events cannot be utilized for
experimental studies in common cases. Therefore, only `detectable'
events should be taken into account in the estimates for a
specific purpose, i.e., events with proper kinematic cuts on $p_T$
and $y$ must be put on precisely in the estimates. Considering
detectors' abilities and to offer experimental references, we try
various cuts accordingly in the estimate of the $B_c$ production.

\begin{figure}
\centering
\hfill\includegraphics[width=0.43\textwidth]{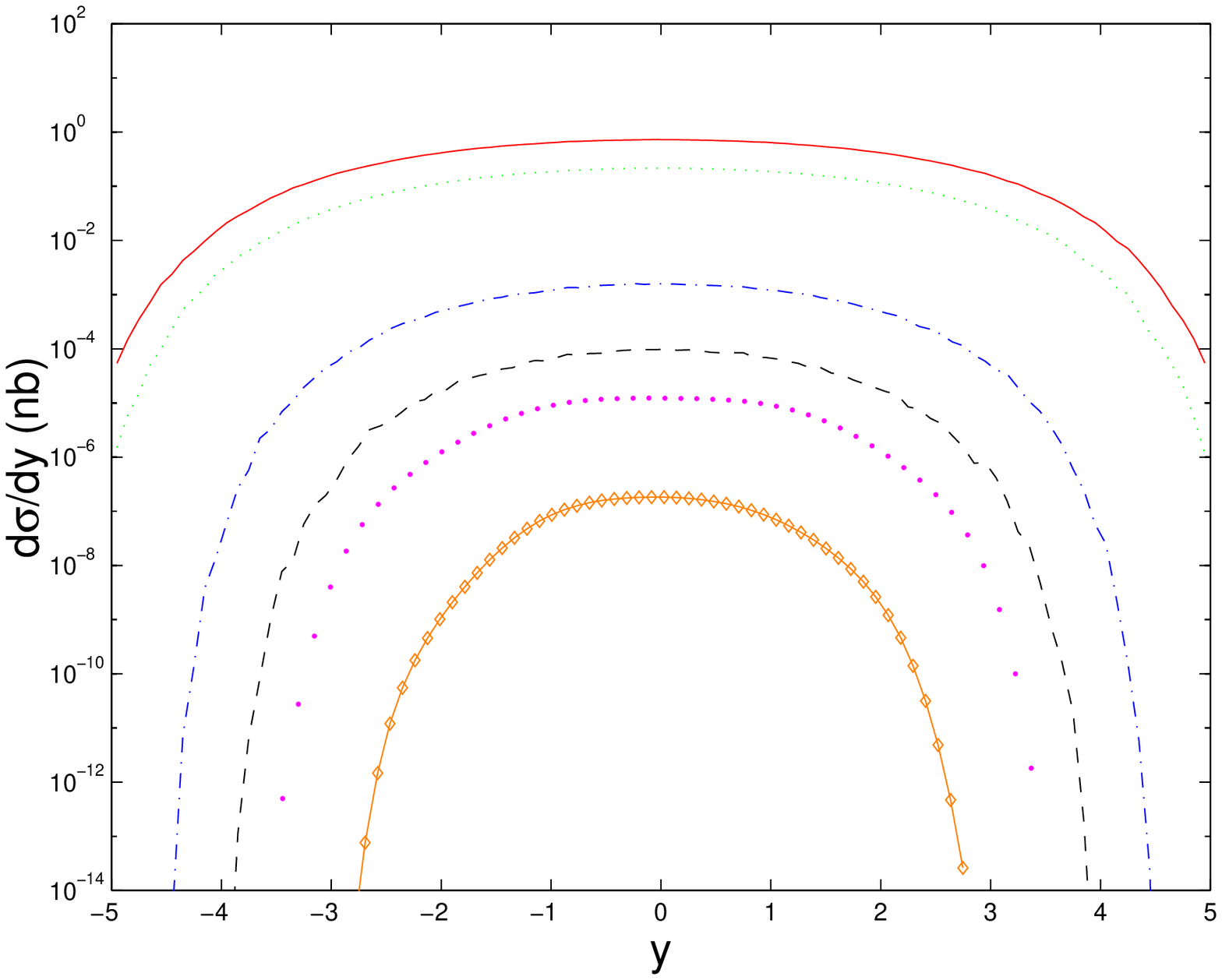}%
\includegraphics[width=0.43\textwidth]{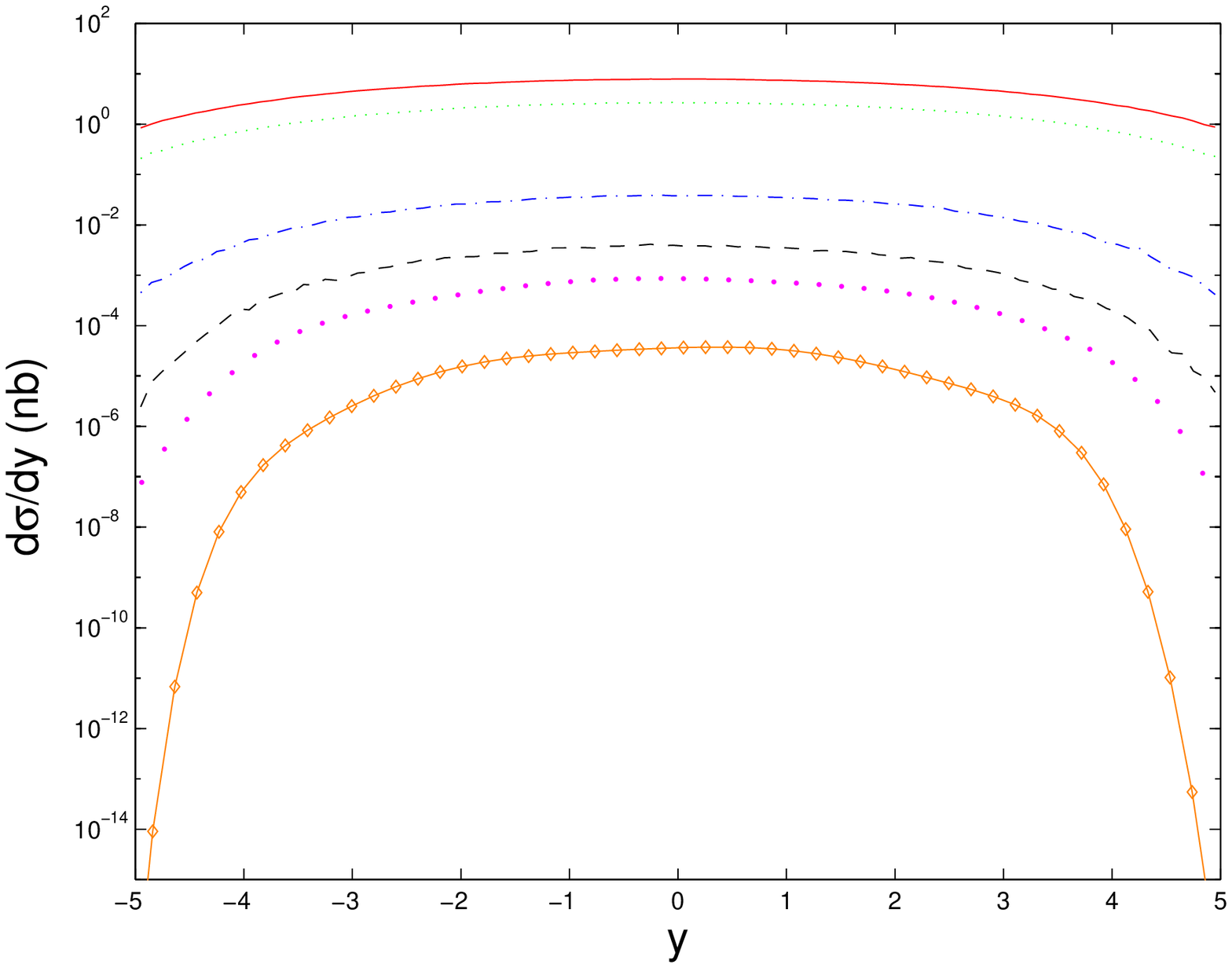}\hspace*{\fill}
\caption{$B_c$ differential distributions versus its $y$ with
various $p_{Tcut}$ in TEVATRON (left diagram) and in LHC (right
diagram). Solid line corresponds to the full production without
$p_{Tcut}$; dashed line to $p_{Tcut}=5.0$ GeV; dash-dot line to
$p_{Tcut}=20.0$ GeV; the dashed line to $p_{Tcut}=35.0$ GeV; the
big dotted line to $P_{Tcut}=50.0$ GeV and the solid line with
diamonds to $p_{Tcut}=100$ GeV.} \label{tptrap}
\end{figure}

First, we study the distributions of $p_T$ and $\sqrt{\hat{s}}$
for the $B_c$ meson with various rapidity cuts $y_{cut}$ (here we
mean that only the $B_c$ events with $|y|\leq y_{cut}$ are taken
into account), and as an example only the pseudo-scalar $B_c$
meson is taken into account. Note that the momenta and energies of
the $b$, $\bar{c}$ and $B_c$ meson are measurable experimentally
(a complete measurement of the events), hence the C.M. energy
$\sqrt{\hat{s}}$ of the subprocess is measurable, i.e., in fact
the distribution of $\sqrt{\hat{s}}$ is measurable experimentally.
Take into account the abilities in measuring rapidity of $B_c$ for
the detectors CDF, D0 and BTeV at TEVATRON, and ATLAS, CMS and
LHC-B at LHC, various possible rapidity cuts, $y_{cut}\sim 1.5$ or
higher, are tried. To project out the cut effects, here we fix to
use CTEQ5L for gluon distribution function, LO running $\alpha_s$
and Type A energy scale ($Q^2=\hat{s}^2/4$) to carry out the
study. The results for the distributions of $p_T$ and $\hat{s}$
with various $y$-cuts are shown a similar behavior, thus we do not
plot the curve here, but note that the dependence of the
differential distributions on $y_{cut}$ and $p_{Tcut}$ for LHC is
stronger than that for TEVATRON. This is because that at TEVATRON,
all the distributions with sizable rapidity cover a smaller region
in $y$. The correlations between $p_T$ and $y$ are interesting, so
we plot the $y$-distributions with various $p_T$-cuts over a wide
range $p_{Tcut}: 5.0\sim 100$ GeV in FIG.\,\ref{tptrap}. From the
figure (FIG.\,\ref{tptrap}), the dependence of the differential
distributions on rapidity $y$ with different $p_{Tcut}$ at LHC
exhibits a broader profile than that at TEVATRON.

The $p_T$-distributions of the production vary with $y_{cut}$
mainly due to the fact that as $p_T$ increases, the dependence of
the distribution on $y$ becomes smaller as the value of $y_{cut}$
becomes less important. When $p_T$ increases to sufficient large
value, the `border' of the differential distributions (see FIG.\,
\ref{tptrap}) becomes smaller than $y_{cut}$ and the $p_T$
differential distributions with and without $y$-cut is coincide.

To analyze the quantitative difference of the differential
distributions with regard to $p_{Tcut}$ and $y_{cut}$, we
introduce a ratio for the integrated hadronic cross sections:
\begin{equation}
R_{p_{Tcut}}=\left(\frac{\sigma_{y_{cut}}}{\sigma_{0}}
\right)_{p_{Tcut}}\,
\end{equation}
where $\sigma_{y_{cut}}$ and $\sigma_{0}$ are the hadronic cross
section with and without $y_{cut}$ respectively. The ratio
$R_{p_{Tcut}}$ varies with $p_{Tcut}$ and $y_{cut}$, and its
values are given in TABLE\, \ref{tevcut}.

\begin{table}
\begin{center}
\caption{Values of the ratio $R_{p_{Tcut}}$ (see definition in
text) for the hadronic production of pseudo-scalar $B_c$ meson in
TEVATRON and LHC.} \vskip 0.6cm
\begin{tabular}{|c||c|c|c||c|c|c||c|c|c||c|c|c||c|c|c|}
\hline\hline $p_{Tcut}$ & \multicolumn{3}{|c||}{$0.0$ GeV}&
\multicolumn{3}{|c||}{$5$ GeV}& \multicolumn{3}{|c||}{$20$ GeV}&
\multicolumn{3}{|c||}{$35$ GeV}
 & \multicolumn{3}{|c|}{$50$ GeV}\\
\hline $y_{cut}$& 1.0 & 1.5 & 2.0 & 1.0 &
1.5 & 2.0 & 1.0 & 1.5 & 2.0& 1.0 & 1.5 & 2.0& 1.0 & 1.5 & 2.0\\
\hline\hline $R_{p_{Tcut}}$ (TEVATRON) & 0.45 & 0.64  & 0.79 &
0.46 & 0.65 & 0.80 & 0.57 & 0.77 &
0.91 & 0.65 & 0.85 & 0.95 & 0.70 & 0.90 & 0.98\\
\hline $R_{p_{Tcut}}$ (LHC) & 0.31 & 0.46  &  0.59 & 0.32 & 0.47 &
0.60 & 0.38 & 0.54 &
0.69 & 0.42 & 0.60 & 0.74 & 0.45 & 0.64 & 0.79\\
\hline\hline
\end{tabular}
\label{tevcut}
\end{center}
\end{table}

From TABLE\, \ref{tevcut} we may see that, for a fixed $y_{cut}$,
the value of $R_{p_{Tcut}}$ becomes larger with increasing
$p_{Tcut}$. It is understandable that the differential
distributions versus the rapidity $y$ decrease with the increment
of $p_{T}$, so the contributions to the hadronic cross section
surviving after the cut, i.e. $(|y|\leq y_{cut})$, increase with
the increment of $p_{Tcut}$.

As for the $B_c $ meson production, only in high energy hadronic
colliders can numerous $B_c$ mesons be produced. The cross section
of $B_c$ meson production is quite sensitive to the C.M. energy of
the collider. Since RUN-II of TEVATRON is running now at a
slightly higher energy than that $1.8$ TeV (of RUN-I), we take
TEVATRON as an example to study the increase caused by slight
changes in the C.M. energy. The C.M. energy for RUN-II is designed
to be $2.0$ TeV, so here we take energy values, $1.8$, $1.9$,
$1.96$ and $2.0$ TeV to compute the cross sections. The obtained
results for the values of the integrated cross-section with
suitable cuts are shown in TABLE\, \ref{tevatron}. From TABLE
\ref{tevatron}, we see that when the C.M. energy of TEVATRON
increases from $1.8$ to $2.0$ TeV, the cross-section increases by
about $20\%$.

\begin{table}
\begin{center}
\caption{The integrated hadronic cross section for TEVATRON at
different C.M. energies. The gluon distribution is chosen from
CTEQ5L and the characteristic energy scale of the production is
chosen as Type A, i.e. $Q^2=\hat{s}/4$. In addition, a cut for
transverse momentum $p_T$ ($p_T < 5$ GeV) and a cut for rapidity
$y$ ($|y|> 1.5$) have been imposed.} \vskip 0.6cm
\begin{tabular}{|c||c|c|c|c|}
\hline \hline C.M. energy & ~~1.8(TeV)~~ & ~~1.9(TeV)~~
& ~~1.96(TeV)~~ & ~~2.0(TeV)~~ \\
\hline\hline
$B_c[^{1}S_0]$   & 0.40 & 0.44 & 0.46 & 0.47\\
\hline
$B_c^*[^{3}S_1]$ & 1.00 & 1.09 & 1.14 & 1.18\\
\hline\hline
\end{tabular}
\label{tevatron}
\end{center}
\end{table}

\section{summary}

In this paper we have presented quantitative studies on the
uncertainties in estimates of the $B_c$ meson hadronic production.
The uncertainties have been examined in turn by `factorizing'
their origins. The computations are based on the generator
BCVEGPY\cite{wuxglun} mainly. The investigated quantitatively
uncertainties involve those due to various versions of PDFs given
by various groups, the variations of the parameters relevant to
the potential model, the strong coupling $\alpha_s$ relevant to
its running and the characteristic energy scale $Q^2$ of the
process where the QCD factorization is carried out and {\it etc.}.
We find that the characteristic energy scale $Q^2$ of the process
is the source which causes the greatest uncertainty for the LO QCD
estimates. We have also shown the differences between LHC and
TEVATRON for various observables with reasonable kinematic cuts,
such as the cuts on the $B_c$ meson transverse momentum $p_{Tcut}$
and rapidity $y_{cut}$. We also point out that at TEVATRON from
RUN-I to RUN-II due to the increase of the collision C.M. energy,
the cross-section of the $B_c$ production increase by about
$20\%$. In view of the fact that TEVATRON is running while LHC is
still under constructing, based on our studies we can clearly see
that experimental $B_c$ studies at the two colliders are
stimulative and complimentary. As for the topic to study of
$B_s-\bar{B_s}$ mixing and $CP$ violations in $B_s$ meson decays
at high energy hadronic colliders with tagged $B_s$ mesons
produced through the $B_c$ decays, since LHC has much higher C.M.
energy, so the $B_c$-production cross-section is higher than one
order of magnitude at TEVATRON, and LHC has much higher luminosity
than at TEVATRON, it seems that the particularly interesting above
topic may be more accessible and fruitful at LHC than that at
TEVATRON RUN-II, even RUN-III.

\vspace{18mm}

\noindent {\large\bf Acknowledgement} The authors would like to
thank Yu-Qi Chen and Guo-Ming Chen for discussions, and to thank
Vaia Papadimitriou for the suggestion to compute the production at
various C.M. energies for TEVATRON. This work was supported in
part by Nature Science Foundation of China (NSFC).
\\


\begin{thebibliography}{s2}

\bibitem{lep} DELPHI Collaboration, P. Abreu {\em et al.} Phys.
Lett. B {398}, 207 (1997); ALEPH Collaboration, R. Barate {\em et
al.} Phys. Lett. B {402}, 213 (1997); OPAL Collaboration, K.
Ackerstaff {\em et al.} Phys. Lett. B {420}, 157 (1998).

\bibitem{CDF} CDF Collaboraten, F. Abe, {\em et al.}, Phys. Rev.
Lett. {\bf 77}, 5176 (1996).

\bibitem{CDF1} CDF Collaboraten, F. Abe, {\em et al.}, Phys. Rev. Lett.
{\bf 81}, 2432 (1998); Phys. Rev. {\bf D58}, 112004 (1998).

\bibitem{B-work} K. Anikeev, {\em et al.}, hep-ph/0201071.

\bibitem{mord} Chao-Hsi Chang, {\it Proceedings of the XXXVIIth
RENCONITRES DE MORIOND} Les Arcs, Savoie, France, {\bf 2002 QCD
and High Energy Hadronic Interactions}, p-27,  hep-ph/20205112.

\bibitem{qigg} C. Quigg, {\it Proceedings of the Workshop on B
Physics at Hadron Accelerators}, Snowmass (CO) USA, 1993, Eds. P.
McBride and C.S. Mishra.

\bibitem{prod0} Chao-Hsi Chang and Yu-Qi Chen, Phys. Rev. D
{\bf 46}, 3854 (1992); Erratum Phys. Rev. {\bf 50}, 6013 (1994).

\bibitem{prod1} Chao-Hsi Chang and
Yu-Qi Chen, Phys. Rev. D {\bf 48}, 4086 (1993).

\bibitem{prod} E.Braaten, K. Cheung and T.C. Yuan, Phys. Rev. D {\bf 48}, 4230
(1993); E. Braaten, K. Cheung and T.C. Yuan, Phys. Rev. D {\bf
48}, R5049 (1993).

\bibitem{prod2} Chao-Hsi Chang,
Yu-Qi Chen, Guo-Ping Han and Hung-Tao Jiang, Phys. Lett. B {\bf
364}, 78 (1995); Chao-Hsi Chang, Yu-Qi Chen and R. J. Oakes, Phys.
Rev. D {\bf 54}, 4344 (1996); K. Kolodziej, A. Leike and R.
R\"uckl, Phys. Lett. B {\bf 355}, 337 (1995).

\bibitem{prod3}  A.V.
Berezhnoy, V.V. Kiselev, A.K. Likhoded, Z. Phys. A {\bf 356}, 79
(1996); S.P. Baranov, Phys. Rev. D {\bf 56} 3046, (1997).

\bibitem{prod4} K. Cheung, Phys. Lett. B {\bf 472}, 408 (2000).

\bibitem{avaa}  A.V. Berezhnoy, V.V. Kiselev, A.K. Likhoded and A.I.
Onishchenko, Phys. Atom. Nucl. {\bf 60}, 1729 (1997);
hep-ph/9703341.

\bibitem{spec}Yu-Qi Chen and Yu-Ping Kuang, Phys. Rev. D {\bf 46},
1165 (1992); E. Eichten, C. Quigg, Phys. Rev. D {\bf 49} 5845
(1994); Phys. Rev. D {\bf 52}, 1726 (1995); S.S. Gershtein, V.V.
Kiselev, A.K. Likhoded and A.V. Tkabladze, Phys. Rev. D {\bf 51},
3613 (1995).

\bibitem{latt} C. T. H. Davies, et al., Phys. Lett. B {\bf 382},
131 (1996).

\bibitem{chen} Chao-Hsi Chang, Yu-Qi Chen, Phys. Rev. {\bf D49}, 3399
(1994); Chao-Hsi Chang and Yu-Qi Chen, Commun. Theor. Phys. {\bf
23} (1995) 451.

\bibitem{dec} A. Abd El-Hady, J.H. Munoz and J.P. Vary; Phys. Rev. D{\bf 62}
014014 (2000).

\bibitem{dec1} N. Isgur, D. Scora, B. Grinstein and M. Wise, Phys. Rev.
D {\bf 39}, 799 (1989);  M. Lusignoli and M. Masetti, Z. Phys. C
{\bf 51}, 549 (1991); D. Scora and N. Isgur, Phys. Rev. D {\bf
52}, 2783 (1995); Dongsheng Du, G.-R. Lu and Y.-D. Yang, Phys.
Lett. B {\bf 387}, 187 (1996); Dongsheng Du, {\em et al.}, Phys.
Lett. B {\bf 414}, 130(1997); Jia-Fu Liu and Kuang-Ta Chao, Phys.
Rev. D {\bf 56} 4133, (1997); P. Colangelo and F.De Fazio, Phys.
Rev. D {\bf 61} 034012 (2000); V.V. Kiselev, A.E. Kovalsky and
A.K. Likhoded, Nucl. Phys. B {\bf 585} 353 (2000); V.V. Kiselev,
A.K. Likhoded and A.I. Onishchenko, Nucl. Phys. B {\bf 569} 473,
(2000); M.A. Nobes and R.M. Woloshyn, J. Phys. G {\bf 26} 1079,
(2001).

\bibitem{gem} M.A. Ivanov, J.G. K\"orner and O.N. Pakhomova, Phys. Lett.
B {\bf 555}, 189 (2003); A. Faessler et al, Eur. Phys. J. direct C
{\bf 4}, 18 (2002); M.A. Ivanov, J.G. K\"orner and P. Santorelli,
Phys. Rev. D {\bf 63}, 074010 (2001).

\bibitem{life} Chao-Hsi Chang, Shao-Long Chen, Tai-Fu Feng and Xue-Qian
Li, Phys. Rev. D {\bf 64}, 014003 (2001); Commun. Theor. Phys.
{\bf 35}, 51 (2001).

\bibitem{MG} M. Beneke and G. Buchalla, Phys. Rev. D {\bf 53}, 4991 (1996).

\bibitem{CCWZ}Chao-Hsi Chang, J.-P. Cheng and C.-D. L\"u, Phys. Lett.
{\bf B 425}, 166 (1998); P. Colangelo and F. De Fazio, Mod. Phys.
Lett. {\bf A 14}, 2303 (1999); Chao-Hsi Chang, Cai-Dian L\"{u},
Guo-Li Wang and Hong-Shi Zong, Phys. Rev. {\bf D60} 114013, 1999;
Chao-Hsi Chang, Yu-Qi Chen, Guo-Li Wang and Hong-Shi Zong, Phys.
Rev. D {\bf 65}, 014017 (2001); Commun. Theor. Phys. {\bf 35}, 395
(2001); Chao-Hsi Chang, Anjan K. Giri, Rukmani Mohanta and Guo-Li
Wang, J. Phys. G {\bf 28}, 1403, (2002), hep-ph/0204279; Xing-Gang
Wu, Chao-Hsi Chang, Yu-Qi Chen and Zheng-Yun Fang, Phys. Rev. D,
{\bf 67}, 094001 (2003), hep-ph/0209125.

\bibitem{wuxglun} Chao-Hsi Chang, Chafik Driouich, Paula Eerola and Xing-Gang Wu,
Comput.Phys.Commun. {\bf 159}, 192(2004); hep-ph/0309120.

\bibitem{pythia} T. Sjostrand, Comput. Phys. Commun. {\bf 82} (1994)
74.

\bibitem{nrqcd} Geoffrey T. Bodwin, Eric Braaten and G. Peter
Lepage, Phys. Rev. D, {\bf 51}, 1125 (1995); Erratum Phys. Rev. D,
{\bf 55}, 5853 (1997).

\bibitem{5lcteq} H.L. Lai, et al., Eur. Phys. J.{\bf C12}, 375(2000).

\bibitem{6lcteq} H.L. Lai, et al., hep-ph/0201195.

\bibitem{98lgrv} M. Glueck, E. Reya, A. Vogt, Eur. Phys. J.{\bf C5},
461(1998).

\bibitem{2001lmrst} A.D. Martin, R.G. Roberts, W.J. Stirling and R.S. Thorne,
Eur. Phys. J. {\bf C23}, 73(2002).

\bibitem{changwu} Chao-Hsi Chang and Xing-Gang Wu, {\it The mechanism of
quark-antiquark annihilation in hadronic production of the meson
$B_c$}, in preparation.

\end{thebibliography}
\end{document}